\documentclass[reprint,
superscriptaddress,
amsmath,amssymb,
aps,prb,nobibnotes,
noeprint,
floatfix,
]{revtex4-2}

\usepackage[pdftex]{graphicx}
\usepackage{bm}
\usepackage{amsmath}
\usepackage{textcomp}
\usepackage{xr-hyper}
\usepackage{hyperref}
\usepackage[all]{hypcap}
\usepackage{xcolor}
\usepackage{placeins}
\usepackage{enumitem}
\usepackage[version=4]{mhchem}
\allowdisplaybreaks

\makeatletter
\newcommand*{\balancecolsandclearpage}{
  \close@column@grid
  \cleardoublepage
  \twocolumngrid
}
\makeatother

\begin{document}

\title{Effect of pressure on the magnetic properties of \texorpdfstring{(Co$_{0.5}$Fe$_{0.5}$)$_5$GeTe$_2$}{(CoFe)5GeTe2}}
\makeatletter
\let\newtitle\@title
\let\newauthor\@author
\let\newdate\@date
\makeatother

\author{Tam\'as Prok}
\affiliation{Department of Physics, Institute of Physics, Budapest University of Technology and Economics, M\H{u}egyetem rkp.\ 3., H-1111 Budapest, Hungary}
\affiliation{MTA-BME Correlated van der Waals Structures Momentum Research Group, M\H{u}egyetem rkp.\ 3., H-1111 Budapest, Hungary}
\affiliation{Zernike Institute for Advanced Materials, University of Groningen, Groningen, The Netherlands}

\author{Zolt\'an Kov\'acs-Krausz}
\affiliation{Department of Physics, Institute of Physics, Budapest University of Technology and Economics, M\H{u}egyetem rkp.\ 3., H-1111 Budapest, Hungary}
\affiliation{MTA-BME Correlated van der Waals Structures Momentum Research Group, M\H{u}egyetem rkp.\ 3., H-1111 Budapest, Hungary}

\author{Harvey Stanfield}
\affiliation{Department of Physics and Astronomy, University of Manchester, Manchester M13 9PL, United Kingdom}

\author{Bing Zhao}
\affiliation{Department of Microtechnology and Nanoscience, Chalmers University of Technology, SE-41296, G\"oteborg, Sweden}

\author{Michael Leon\'ard Morgan}
\affiliation{Department of Physics, Institute of Physics, Budapest University of Technology and Economics, M\H{u}egyetem rkp.\ 3., H-1111 Budapest, Hungary}

\author{B\'alint F\"ul\"op}
\affiliation{Department of Physics, Institute of Physics, Budapest University of Technology and Economics, M\H{u}egyetem rkp.\ 3., H-1111 Budapest, Hungary}

\author{Marcos H.\ D.\ Guimar\~aes}
\affiliation{Zernike Institute for Advanced Materials, University of Groningen, Groningen, The Netherlands}

\author{Ivan J.\ Vera-Marun}
\affiliation{Department of Physics and Astronomy, University of Manchester, Manchester M13 9PL, United Kingdom}

\author{Saroj Prasad Dash}
\affiliation{Department of Microtechnology and Nanoscience, Chalmers University of Technology, SE-41296, G\"oteborg, Sweden}

\author{Szabolcs Csonka}
\affiliation{Department of Physics, Institute of Physics, Budapest University of Technology and Economics, M\H{u}egyetem rkp.\ 3., H-1111 Budapest, Hungary}
\affiliation{MTA-BME Superconducting Nanoelectronics Momentum Research Group, M\H{u}egyetem rkp.\ 3., H-1111 Budapest, Hungary}
\affiliation{HUN-REN Centre for Energy Research, Institute of Technical Physics and Materials Science, Konkoly Thege Miklós út 29-33, 1121 Budapest, Hungary}

\author{P\'eter Makk}
\email{makk.peter@ttk.bme.hu}
\affiliation{Department of Physics, Institute of Physics, Budapest University of Technology and Economics, M\H{u}egyetem rkp.\ 3., H-1111 Budapest, Hungary}
\affiliation{MTA-BME Correlated van der Waals Structures Momentum Research Group, M\H{u}egyetem rkp.\ 3., H-1111 Budapest, Hungary}

\author{Endre T\'ov\'ari}
\email{tovari.endre@ttk.bme.hu}
\affiliation{Department of Physics, Institute of Physics, Budapest University of Technology and Economics, M\H{u}egyetem rkp.\ 3., H-1111 Budapest, Hungary}
\affiliation{MTA-BME Correlated van der Waals Structures Momentum Research Group, M\H{u}egyetem rkp.\ 3., H-1111 Budapest, Hungary}

\date{\today}

\begin{abstract}
Cobalt-doped Fe$_5$GeTe$_2$ possesses a rich magnetic phase diagram as a function of Co concentration. The nature of magnetic order in (Co$_{0.5}$Fe$_{0.5}$)$_5$GeTe$_2$ is especially interesting, as it has been shown to exhibit ferromagnetic order, A-type antiferromagnetic (AFM) order, or potentially both at the same time. Here we present magnetoresistance measurements on antiferromagnetic (Co$_{0.5}$Fe$_{0.5}$)$_5$GeTe$_2$ at a series of pressures and extract the anisotropy and interlayer exchange fields using the two-sublattice model. We show a 50\% increase of the interlayer exchange at 2\,GPa, highlighting the sensitivity of magnetic properties to interlayer distance. In addition, we find that the sharp hysteretic transitions observed within the AFM state can be qualitatively described by a linear chain model, which suggests an even-odd effect as a function of layer number instead of a coexisting ferromagnetic phase.
\end{abstract}

\maketitle
\counterwithin*{equation}{part}
\stepcounter{part}
\renewcommand{\theequation}{\arabic{equation}}
\section{Introduction}\label{sec:intro}

For spintronics applications, antiferromagnetic (AFM) materials possess multiple advantages compared to ferromagnets (FM). They exhibit negligible stray fields, are not sensitive to magnetic perturbations, and their internal spin dynamics are significantly faster\cite{Baltz2018, 2024Din}. Moreover, they demonstrate  potential in memory applications due to a large magnetoresistance, without requiring the fabrication of a multilayer tunnel junction device\cite{Song2018, Klein2018}. Two-dimensional (2D) magnets are especially intriguing as they show a wide variety of magnetic ordering down to the monolayer thickness, and may be controlled by electric fields, current, or other means\cite{Ningrum2020, Elahi2022, Mi2023, Jia2025}. Due to their novel properties, atomic thickness, and impurity-free interfaces, 2D heterostructures built of magnetic and nonmagnetic layers offer new opportunities for the design of future spintronic devices\cite{Sierra2021, Yang2022, Song2023, Roche2024, Thapa2025, Zhao2026arxiv}. 

Among 2D magnets, Fe$_3$GeTe$_2$ and Fe$_3$GaTe$_2$ have been shown to exhibit a moderate stability in air, a Curie temperature $T_C$ approaching or exceeding room temperature, and they have a strong perpendicular magnetic anisotropy (PMA) relative to the molecular plane\cite{Fei2018, Deng2018FGT, Zhang2022_FGaT}. Their magnetic properties are highly sensitive to the composition, for example, increasing the Fe content can increase $T_C$ on the order of a hundred Kelvins as shown in Fe$_5$GeTe$_2$\cite{May2019_F5GT, Zhao2023AM}. Moreover, by substituting Fe with Co in Fe$_5$GeTe$_2$, the magnetic order can change drastically. Defining the Co concentration $x$ as (Co$_x$Fe$_{1-x}$)$_5$GeTe$_2$ (CFGT for short), for approximately $x=0.1-0.3$ the FM easy direction is in the $ab$ plane\cite{Tian2020APL, May2020PRM, Ngaloy2024}. In contrast, in the $x=0.4-0.5$ range, CFGT is an A-type antiferromagnet, i.e.\ magnetic moments in each molecular layer are ferromagnetically coupled, but the magnetization of neighbouring layers is opposite, as depicted in Fig.~\ref{fig1}a. For these Co concentrations, PMA is recovered and the N\'eel temperature $T_N$ ranges between approximately 300 and 430\,K\cite{Tian2020APL, May2020PRM, Lu2024NanoLett, Zhao2025aAdvMat}. However, a FM state has also been observed at $x=0.5$ \cite{Zhang2022PRM, Zhang2024AdvMat}, and it is also predicted to be present at $x=0.6$ \cite{May2020PRM}. Moreover, coexisting AFM and FM order has been reported based on hysteretic features inside the AFM phase\cite{Lu2024NanoLett, Zhao2025aAdvMat}. The changes with increasing $x$ have been tied to transitions from ABC to AA to AA' stacking\cite{May2020PRM, Zhang2022PRM}, although it is unclear why both FM and AFM states are possible around 50\% Co substitution of Fe. 

Here we show that the magnetic characteristics of CFGT can be actively tuned by another control knob: the interlayer distance via pressure\cite{Fulop2021}. Hydrostatic pressure has been shown to induce changes in stacking and magnetic order\cite{Li2019NatMat, Song2019NatMat}. We performed magnetoresistance measurements on multiple (Co$_{0.5}$Fe$_{0.5}$)$_5$GeTe$_2$ samples and monitored their magnetic properties via the anomalous Hall effect (AHE)\cite{Nagaosa2010}. We have found that the antiferromagnetic interlayer coupling is enhanced by 50\% by a pressure $p$ of 2\,GPa. In addition, we propose that the hysteresis loops observed within the AFM phase can be explained by the behaviour of the surface layers instead of the presence of a bulk FM order.

\section{Methods}\label{sec:methods}

Single crystals of $\ce{(Co_{0.5}Fe_{0.5})_5GeTe_2}$, grown by chemical vapor transport (CVT), were obtained from HQ Graphene. A scanning transmission electron microscope image is shown in Fig.~\ref{fig1}a, further characterization can be found in Ref.~\citenum{Zhao2025aAdvMat}. We prepared 10 to 90\,nm thick flakes by exfoliation onto $\ce{SiO_2}$ substrates. Their thickness was measured by atomic force microscopy. Chromium/gold leads were deposited via standard electron beam lithography after cleaning the surface of the contact areas by argon-milling.

Figure~\ref{fig1}b shows a photo of sample A, an approximately 90-nm-thick flake, and also illustrates the measurement geometry. We studied the Hall resistance $R_{xy}$ of CFGT samples as a function of an applied out-of-plane magnetic field $H$ at a series of temperatures $T$ and hydrostatic pressures $p$ up to 2\,GPa. To apply pressure the samples were placed into a pressure cell which was filled with kerosene\cite{Fulop2021} during all of the measurements. The electrical measurements were performed in a variable temperature insert (VTI), using lock-in amplifiers and differential amplifiers. The current used for the resistance measurements was generally $I_\text{rms} = 10\,\mathrm{\mu A}$ at $f = 132.43\,\mathrm{Hz}$.

\section{Results}\label{sec:results}

\begin{figure}[!t]
\begin{center}
	\includegraphics[width=\columnwidth]{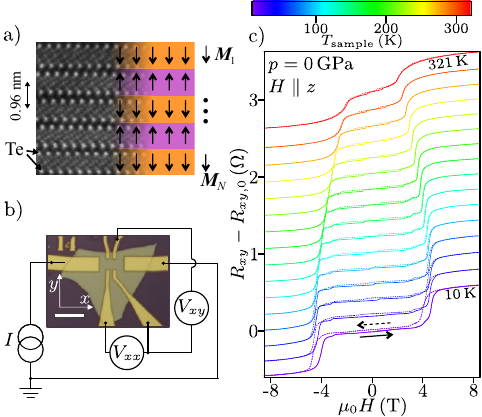}
	\caption{
	a) Scanning transmission electron microscope image of a CFGT crystal. The brightest spots are Te atoms. The alternating magnetization in neighboring layers in the AFM state is illustrated. The thickness of a molecular layer and its magnetization are indicated.  
    b) Optical image of 90-nm-thick sample A and a diagram of the measurement setup. The scale bar is $10\,\mathrm{\mu m}$.
    c) The Hall resistance of the sample at a series of selected temperatures. Solid (dotted) lines indicate data for up (down) sweeps of the field. A zero-field value $ R_{xy,0} $, which is a result of the mixing of a longitudinal voltage drop, has been removed. The curves are plotted with offsets that are proportional to $T$. 
	}\label{fig1}\end{center}
\end{figure}

The Hall resistance of sample A at ambient pressure is shown in Fig.~\ref{fig1}c from approximately 10\,K up to above room temperature, after removal of a zero-field offset $R_{xy,0}$. All $R_{xy}$ curves show step-like features that are symmetrical around $H=0$. These are characteristic of an antiferromagnet with an approximately out-of-plane easy axis\cite{Coey2010, Baltz2018, Tian2020APL, Lu2024NanoLett, Zhao2025aAdvMat} where the behavior of the out-of-plane component of the magnetization $\bm{M}$ is reflected in $R_{xy}$ via the AHE. A step during increasing $|H|$ corresponds to a first order phase transition from the AFM state towards the fully polarized FM phase. As we show further below, this is a spin flop (SF) transition in which the AFM state switches to a canted AFM state, and denote the corresponding field by $H_\mathrm{SF}$. At high fields, $R_{xy}$ appears to saturate, over an approximately linear background due to the conventional Hall effect (CHE). At most temperatures, the hysteresis between up- and down-sweeps of $H$ is negligible at the SF transition, and only becomes clearer below $\sim 50$\,K.

Another interesting feature of the data on sample A is a pair of small rectangular hysteresis loops at most $T$. As illustrated in Fig.~\ref{fig1}c and magnified in Fig.~\ref{fig4}b, they are observable within the AFM phase ($|H|<H_\mathrm{SF}$) even above room temperature. Each loop can be characterized by a pair of coercive fields with the same sign, and therefore have a negligible remanence at zero field, similarly to certain samples in Ref.~\citenum{Lu2024NanoLett}. The exception is below 100\,K, where the shift of the coercive fields leads to an overlap of the hysteresis loops and a finite remanence.

\begin{figure*}[!t]
\begin{center}
	\includegraphics[width=\textwidth]{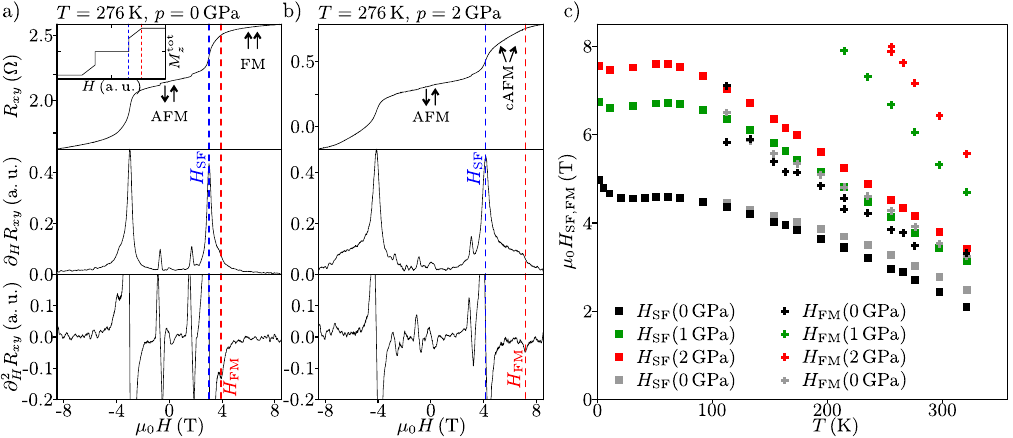}
	\caption{
	a-b) Examples of the raw Hall data and its derivatives on sample A at 276\,K as a function of increasing $H$ at $p = 0$ and 2\,GPa, respectively. The vertical blue and red lines mark the $H_\mathrm{SF}$ and $H_\mathrm{FM}$ fields, respectively. Arrows illustrate the AFM, canted AFM and FM phases. The peaks in the derivatives within the AFM phase are due to the steps of the small hystereses. Inset: simulation of the total out-of-plane magnetization using the two-sublattice model.
    c) The values of the $H_\mathrm{SF}$ (solid squares) and the $H_\mathrm{FM}$ (crosses) fields as a function of $T$ at different pressures.}\label{fig2}\end{center}
\end{figure*}

We extracted $H_\mathrm{SF}$ by identifying sharp peaks in the first derivative $\partial R_{xy} / \partial H$ as shown in Fig.~\ref{fig2}a for an up-sweep of $H$. At low temperature where the SF transition is hysteretic, we averaged the $H_\mathrm{SF}$ of up and down sweeps. The results are plotted in Fig.~\ref{fig2}c as solid black squares as a function of temperature. $\mu_0 H_\mathrm{SF}$ decreases from approximately $5$ to $\sim 2$\,T between 1.5 and 320\,K, which likely reflects the decrease of the absolute layer magnetization with increasing $T$. Under applied pressures up to $p=2$\,GPa, the magnetoresistance curves are qualitatively similar to those taken at $p=0$\,GPa, as illustrated in Fig.~\ref{fig2}b. The obtained $H_\mathrm{SF}$, plotted in Fig.~\ref{fig2}c as a function of temperature for all pressures with solid squares, show similar trends with $T$, while we observe a monotonic increase in $H_\mathrm{SF}$ with $p$. The measurements performed after the release of pressure are shown in grey, and closely match the values before applying pressure (black). All datasets suggest that the Néel temperature is well above 350\,K, in accordance with Ref.~\citenum{Zhao2025aAdvMat}.



We have investigated further CFGT samples and obtained similar results regarding the main features of the $R_{xy}$ curves. Most data shown in the main text were measured on sample A unless stated otherwise. For an overview of all measurements, we refer to the Supplementary Information (SI) Section~1.

\section{Discussion}\label{sec:discussion}

In the following, we show a framework that describes the dominant features of the magnetoresistance of CFGT, and reveal a giant tunability of the model parameters with pressure. Then we turn to the peculiar fine structure of the $R_{xy}$ curves, i.e.\ the small hysteresis loops within the AFM phase, taken on several samples. We demonstrate that the model serves as a good foundation to explain the details of the magnetic behaviour without invoking a simultaneous FM order.

Here we examine the nature of the magnetic phase transition at $H_\mathrm{SF}$. Since $H$ is approximately aligned with the easy axis\cite{Zhao2025aAdvMat}, the step is either a spin flip transition directly from the AFM to the FM phase, or a spin flop transition to an intermediate, canted AFM (cAFM) state where the layer magnetizations are not collinear\cite{Baltz2018}. In the latter case, the net magnetization jumps to a finite value at $H_\mathrm{SF}$, followed by a continuous change during the cAFM phase until the onset of the FM phase, as shown in the inset of Fig.~\ref{fig2}a. This point is denoted by $H_\mathrm{FM}$ and is accompanied by a dip in the second derivative, $\partial^2 R_{xy} / \partial H^2$, while it is absent in the spin flip case. Such a feature is clearly visible in the second derivative of the data as highlighted by the red dashed line in Fig.~\ref{fig2}b, confirming the presence of the cAFM state at 2\,GPa. A dip of the second derivative is also present at 0\,GPa (Fig.~\ref{fig2}a), although it is less distinct, as its distance to $H_\mathrm{SF}$ is four times smaller. These results show that pressure stabilizes the cAFM phase compared to the ambient state. The obtained $H_\mathrm{FM}$ fields are plotted in Fig.~\ref{fig2}c as crosses. We note that we could only extract $H_\mathrm{FM}$ for a limited range in temperature since it increases with decreasing $T$ and goes above our experimental limitations ($\mu_0|H|=8.5$\,T).

\begin{figure}[!th]
\begin{center}
	\includegraphics[width=\columnwidth]{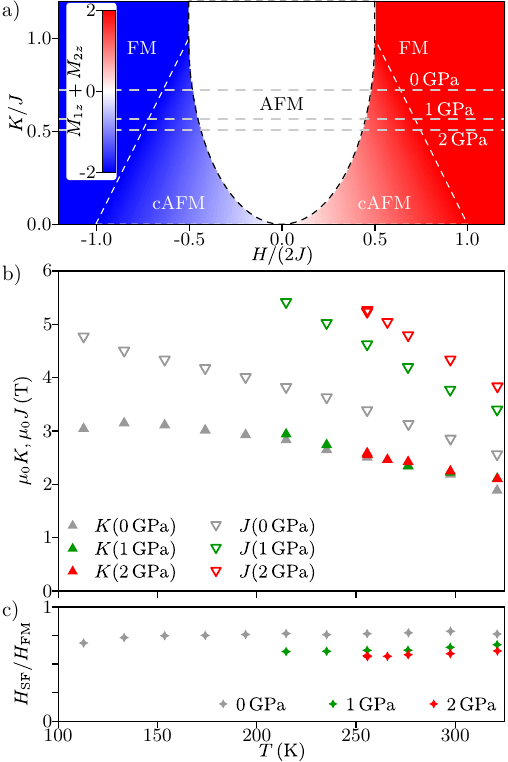}
	\caption{
    a) Phase diagram of the two-sublattice system according to the numerically calculated global minimum of Eq.~\ref{eq:energy}. The black dashed line shows the analytical first-order spin-flop/flip phase transition at $\pm H_\mathrm{SF}$, given by Eq.~\ref{eq:Hsf1} for $K/J<1$ and $\pm J$ for $K/J>1$. The white dashed lines are the second-order phase transition from the cAFM to the FM phase at $\pm H_\mathrm{FM}$, from Eq.~\ref{eq:Hfm1}. Both fields are rescaled by $2J$. The grey dashed lines are the $K/J$ values determined from experimental $H_\mathrm{SF} / H_\mathrm{FM}$ data for sample A at different pressures.
	b) The effective $K$ (solid triangles) and $J$ (empty triangles) values as a function of temperature, extracted using Eq.~\ref{eq:HtransCalc} and experimental $H_\mathrm{SF}, H_\mathrm{FM}$.
    c) The measured $H_\mathrm{SF}/H_\mathrm{FM}$ as a function of temperature.
	}\label{fig3}\end{center}
\end{figure}

In order to gain further insight into the main magnetoresistance features and the role of pressure, we turn to the two-sublattice model of A-type AFMs\cite{Feder1968}. This assumes that each layer has uniform, saturated magnetization and that the surface layers' magnetization is negligible compared to the bulk. This applies to sample A, where the layer number $N$, using a layer thickness of 0.96\,nm (see also Fig.~\ref{fig1}a)\cite{Tian2020APL, Zhang2022PRM}, is around $94$. Therefore, every second layer can be considered equivalent in the AFM phase, and the system can be described as two coupled sublattices. The directions of the sublattice magnetizations $\bm{M}_{1,2}$ can be determined by minimizing the following energy functional:
\begin{equation}\label{eq:energy}
    f = J \bm{M}_1\cdot\bm{M}_2 - \sum_{i=1,2} \left( \frac{K}{2} M_{iz}^2 + \bm{H}\cdot\bm{M}_i \right)
\end{equation}
where the signed bracket is essentially the Stoner-Wohlfarth energy\cite{Aharoni2020}. Here, units of A/m are used for the interlayer exchange $J>0$, the anisotropy $K>0$ and the applied field $\bm{H}$, while $\bm{M}_{1,2}$ are dimensionless unit vectors. We have calculated the phase diagram of the system by identifying the lowest-energy stable solution (global minimum) at any point on the $H/2J,K/J$ plane. It is plotted as a colormap of the $z$ component $M_z$ of the total magnetization ($\bm{M} = \bm{M}_1 + \bm{M}_2$) in Fig.~\ref{fig3}a. The model predicts that if $K<J$, we may observe the AFM, cAFM, then FM phases as $H$ is increased from 0, with distinct transitions at $H_\mathrm{SF}$ and $H_\mathrm{FM}$. We can express the phase transition fields defined above as
\begin{subequations}
	\label{eq:HtransCalc} 
	\begin{eqnarray}
    H_\mathrm{SF} &=& \sqrt{K(2J - K)},
	\label{eq:Hsf1}
	\\
    H_\mathrm{FM} &=& 2J - K. \label{eq:Hfm1}
	\end{eqnarray}
\end{subequations}
We plot the theoretical $H_\mathrm{SF}$ and $H_\mathrm{FM}$ in Fig.~\ref{fig3}a after rescaling with $2J$ as black and white dashed lines, respectively, which clearly delineate the three phases. If $K>J$, the canted state is no longer a global energy minimum, leading to a spin-flip transition at $H_\mathrm{SF}=H_\mathrm{FM}=J$. Based on Eqs.~\ref{eq:HtransCalc}a,b and the experimental values of $H_\mathrm{SF}, H_\mathrm{FM}$ in Fig.~\ref{fig2}c, we calculated the effective temperature-dependent parameters $K(T),J(T)$ and plotted them in Fig.~\ref{fig3}b. In general, the anisotropy $K$ is independent of pressure, while the interlayer exchange $J$ increases by as much as 50\% with $p$, likely due to the decreased distance between the layers as in Ref.~\citenum{Marffy2026}. 

The effective $K$ and $J$ decrease as $T$ is increased, which may be attributed to the reduction of the actual magnetization of the layers. We can separate the latter into an unknown, dimensionless $M(T)$ function\cite{Wang2019} and substitute $M(T)\bm{M}_i$ into Eq.~\ref{eq:energy} in place of $\bm{M}_i$. As a result, we can rewrite Eqs.~\ref{eq:HtransCalc}a,b to describe the observed temperature-dependent transition fields as follows:
\begin{subequations}
	\label{eq:HtransCalc2} 
	\begin{eqnarray}
    H_\mathrm{SF}(T) &=& M(T) \sqrt{K(2J - K)},
	\label{eq:Hsf2}
	\\
    H_\mathrm{FM}(T) &=& M(T) ( 2J - K ). \label{eq:Hfm2}
	\end{eqnarray}
\end{subequations}
Now $K,J$ are independent of temperature. While individually they cannot be determined from experiments, a dimensionless parameter can be defined from the ratio of the spin-flop and saturation fields. In this picture $H_\mathrm{SF} (T)/H_\mathrm{FM}(T) =\sqrt{K/(2J-K)} $ is not expected to depend on the temperature and this is confirmed by the experimental ratio shown in Fig.~\ref{fig3}c. From this data we have calculated the $T$-averaged ratio $K/J = 2/(1+(H_\mathrm{FM}/H_\mathrm{SF})^{2})$ and illustrate its evolution with pressure in the phase diagram in Fig.~\ref{fig3}a with horizontal dashed lines. In short, the increasing width of both the AFM and cAFM phases on the $H$ axis with increasing $p$ can be explained within the two-sublattice model, where the interlayer AFM exchange $J$ is substantially enhanced by pressure.



\begin{figure*}[!t]
\begin{center}
	\includegraphics[width=\textwidth]{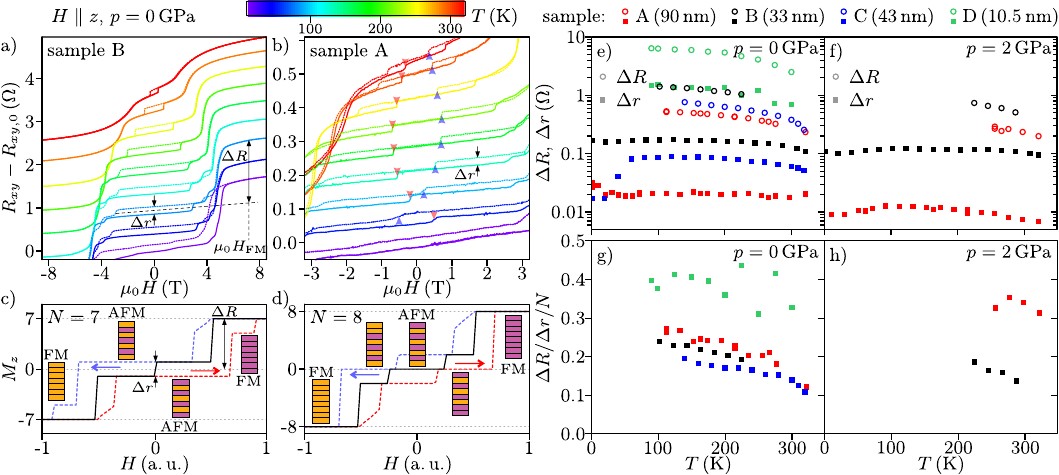}
    \caption{
    a,b) Fine structure of the anomalous Hall resistance in the AFM phase in samples B and A, respectively. The step and small loop heights $\Delta R, \Delta r$ are illustrated. In b), red and blue triangles highlight the coercive fields closest to $H=0$ of the double hysteresis loops. 
    c,d) Simulations of the magnetization $M_z$ for 7 and 8 layers following the global energy minimum of Eq.~\ref{eq:energylin} (black lines), and simulations following the local energy minima for up and down sweeps (red and blue dashed lines). Stacks of rectangles indicate the magnetization $M_{iz}$ of the individual layers in the global energy minimum at each $H$, with purple for positive and orange for negative. e,f) $\Delta R, \Delta r$ in multiple samples at 0 and 2\,GPa, respectively. g,h) The ratio $\Delta R / \Delta r / N$ at 0 and 2\,GPa. 
    } \label{fig4}\end{center}
\end{figure*}

Finally, we discuss the curious hysteresis loops of the magnetoresistance curves in the AFM state of sample A, as well as our other devices, samples B, C and D which were approximately 33\,nm, 43\,nm and 10.3\,nm thick, respectively. While sample A exhibits double hysteresis loops in the AFM phase, in the other samples we find a single hysteresis loop with finite remanence at $H=0$ as shown in Fig.~\ref{fig4}a from sample B. A remanent resistance in the AFM phase has previously been attributed to the coexistence of a ferromagnetic phase at room temperature\cite{Zhao2025aAdvMat}. In Ref.~\citenum{Lu2024NanoLett} a similar conclusion was drawn concerning even-layer samples from the increase of the remanent anomalous Hall resistance below 75\,K. In the same reference, qualitative differences were observed between samples of even and odd layer number regarding the number of steps in $R_{xy}$, which were attributed to the role of surface layers and explained by modeling the magnetization of all layers, i.e. using the linear chain model. 

Here we show that the linear chain model readily explains these observations without the use of a coexisting FM phase. The model\cite{Wang2019,Ye2022} uses the same parameters as Eq.~\ref{eq:energy}, but treats the magnetization $\bm{M}_i$ of each layer (see Fig.~\ref{fig1}a) individually:
\begin{equation}\label{eq:energylin}
    f' = \frac{J}{2} \sum_{i=1}^{N-1} \bm{M}_i\cdot\bm{M}_{i+1} - \sum_{i=1}^{N} \left( \frac{K}{2} M_{iz}^2 + \bm{H}\cdot \bm{M}_i \right),
\end{equation}
where $N$ is the layer number. Each layer is coupled to two adjacent layers except at the surface, therefore $J/2$ is used. Surface layers are coupled to only one layer, which may lead to them switching at a different field than the bulk.

To illustrate the role of surface layers, we have calculated the global minimum of Eq.~\ref{eq:energylin} with $N=7$ and $8$ layers. The results are discussed below, and are representative of odd and even layer systems within the model, with only quantitative differences between different $N$ with the same parity (see SI). The net magnetization $M_z$ in the global minimum for $N=7,8$ is shown by the black lines in Fig.~\ref{fig4}c,d, respectively. In an odd-layer sample, $M_z$ is always finite due to an uncompensated layer. There are two AFM states whose layer configurations are mirrored. In each AFM state, the layer directions $M_{iz}$ are illustrated by colored rectangles with purple and orange for $M_{iz}>0$ and $M_{iz}<0$, respectively, similarly to Fig.~\ref{fig1}a. The system switches between them at $H=0$ so that $M_z$ remains aligned with $H$. This is shown by a step of the black line from $-1$ to $+1$ at $H=0$ (Fig.~\ref{fig4}c). However, in real samples such first order changes are hysteretic in general, because the system can remain in a state as long as it is an energy minimum, or until an external perturbation such as thermal excitations lead to relaxation into a lower-energy state. As a result, a hysteresis loop is expected to form around each predicted step of the black line. The $H=0$ step should expand into a hysteresis loop as if for a ferromagnet, which matches the experiments on samples B (Fig.~\ref{fig4}a) as well as C, D (Figs.~S3, S4 in the SI). 

For $N=8$, the calculated magnetization $M_z$ in the global energy minimum is plotted in Fig.~\ref{fig4}d by the black line. Here the AFM state is well defined around $H=0$ with $M_z=0$. In addition, plateaus of $M_z=\pm2$ can be found between the AFM and FM states, which is characteristic for all even $N \geq 4$. Here, the surface layer that was antiparallel with $H$ in the AFM state is now aligned with $H$ and also its neighbor, as shown by the series of rectangles. Like for odd $N$, in reality a hysteresis loop is expected to form around each step, which is consistent with the double loop structure in sample A, replotted for comparison in Fig.~\ref{fig4}b. 

As temperature decreases, the experimental data becomes more complex, as demonstrated by additional hysteresis loops in Figs.~\ref{fig1}c and~\ref{fig4}a around $\pm H_\mathrm{SF}$. In sample A we also found a remanent AHE resistance, i.e.\ a finite difference at $H=0$ between up- and down sweeps as can be seen in Fig.~\ref{fig4}b, similarly to Ref.~\citenum{Lu2024NanoLett} for even layers. However, the evolution of the small loops with temperature makes it clear that the origin of the remanence, instead of a new FM state, is that the pair of small loops begins to overlap at low $T$, as displayed by the red and blue triangles. As a proof of concept, we have conducted theoretical calculations of the metastable states of the system. Now, instead of the global minimum, we always follow a local energy minimum until it becomes unstable. The red and blue dashed lines in Fig.~\ref{fig4}c,d represent such calculated up- and down field sweeps, leading to a hysteresis with a complex shape. For an even layer number, this predicts a remanent magnetization around $H=0$, leading to a finite AHE which is consistent with our low-$T$ experimental findings in sample A.

The proposition that the switching of surface layers is behind the small hysteresis loops in the AFM phase is further supported by our analysis of the small loop height and the AFM-(cAFM)-FM step height. For this we antisymmetrized the $R_{xy}$ curves\cite{Marffy2026}, and corrected them by a linear background attributed to the conventional Hall effect (CHE), which we determined by fitting the regions near $H=0$. We extracted the magnitude $\Delta R$ of AHE in the FM state at $H_\mathrm{FM}$ as shown in Fig.~\ref{fig4}a, to minimize systematic errors from a potentially nonlinear CHE. This step approximately corresponds to a magnetization change $\Delta M_z^{\Delta R}=N$ in thick samples. $\Delta R$ is shown in Fig.~\ref{fig4}e,f at 0 and 2\,GPa, and its temperature-dependence is comparable to that of $H_\mathrm{SF}(T)$. The height $\Delta r$ of the small loop, estimated far from hysteresis overlaps (see Fig.~\ref{fig4}a,b), corresponds to a $\Delta M_z^{\Delta r}=2$ magnetization change and is also plotted in Fig.~\ref{fig4}e,f. In general, both $\Delta R$ and $\Delta r$ slightly decrease with increasing pressure. They also decrease as the crystal thickness increases, which is consistent with the behaviour of AHE in ferromagnetic sister compounds Fe$_3$GeTe$_2$ and Fe$_3$GaTe$_2$ \cite{Tan2018, Roemer2020, Wang2024b}.

We have calculated the ratio $\Delta R / \Delta r / N$, where the layer number $N$ was estimated from the flake thickness, and plot it in Fig.~\ref{fig4}g,h. As discussed above, it is predicted to be $\Delta M_z^{\Delta R} / \Delta M_z^{\Delta r} / N = 0.5$, but according to our findings it is lower and depends on the temperature. These properties may originate, for example, in a $H$-nonlinear CHE that distorts $\Delta R$, a different $T$-dependence of surface and bulk layer magnetizations (due to the presence or lack of a neighbour), or a different saturation state of the magnetizations as $\Delta r, \Delta R$ are read out at different fields, which the model does not account for. Nevertheless, the observed $\Delta R / \Delta r / N$ is comparable for several samples of various thicknesses, demonstrating a qualitative agreement with the chain model. 

The magnetic properties of CFGT and its sister materials are extremely sensitive to the configuration of magnetic ions\cite{Ershadrad2022, Zhao2025aAdvMat, Tyson2025}. Therefore, the cause of the reduced ratio $\Delta R / \Delta r / N$ may be, in addition to those listed above, due to the effect of surface oxidation\cite{Li2023} on $\Delta r$. As $\Delta r$ is related to AHE in the surface layers, we speculate that an increase in $\Delta r$ may be produced in various ways, from a higher surface current density to having a different local magnetization or even a sign change of the local interlayer coupling $J$ \cite{Kim2019}, which may be detectable via surface mapping techniques\cite{Tschudin2024, Pellet2025}. A stronger surface spin-dependent scattering that enhances AHE is another possibility and may also contribute to the different $T$-dependence of $\Delta R, \Delta r$ in Fig.~\ref{fig4}e,f. Deviations may also be the result of inhomogeneities in the layer magnetization, stacking, etc.








\section{Conclusions}\label{sec:concl}

We have studied the magnetic response in antiferromagnetic $\ce{(Co_{0.5}Fe_{0.5})_5GeTe_2}$ samples between 10 to 90\,nm thickness via the anomalous Hall effect. We have clearly delineated the AFM, cAFM, and FM states and shown that, with pressure, the magnetic transition fields increase substantially and the cAFM state is stabilized. Although the anisotropy is practically unaffected, the interlayer AFM coupling $J$ increases by approximately 50\% when applying 2\,GPa, demonstrating the sensitivity of magnetic parameters to the interlayer distance. In addition, we found that the small hysteresis loops within the AFM phase can be qualitatively explained by the behaviour of the surface layers as predicted by the linear chain model. The analysis of the resistance steps to the FM state compared to those in the small loops is in line with this picture. 

CFGT is the highest-Néel-temperature 2D antiferromagnet discovered to date\cite{Zhao2026arxiv}. We believe that our results shed light on the intriguing magnetic behaviour of the bulk and the surface of CFGT crystals, and will contribute to the design of antiferromagnetic components and interfaces in future spintronic devices.

\section*{Acknowledgements}\label{sec:ack}

This research was supported by the Ministry of Culture and Innovation and the National Research, Development and Innovation Office within the Quantum Information National Laboratory of Hungary (Grant No.\ 2022-2.1.1-NL-2022-00004 and UNKP-23-4-I-BME-36), by OTKA grant No.\ K134437 and the NKKP STARTING grant No.\ 150232. We acknowledge funding from the 2DSOTECH FlagERA network, the 2DSPIN-TECH Flagship project, the Alexander von Humboldt Foundation, the European Research Council ERC project Twistrain, the TRILMAX Horizon Europe consortium (Grant No.\ 101159646), and COST Action CA 21144 superQUMAP. M.\ H.\ D.\ G. acknowledges funding from the Zernike Institute for Advanced Materials and the European Union, through grant ERC StG 2D-OPTOSPIN, Grant No.\ 101076932. H.\ S. and I.\ J.\ V.\ M. in Horizon Europe Project "2D Heterostructure Non-volatile Spin Memory Technology" (2DSPIN-TECH) are supported by UKRI grant number [10101734] (The University of Manchester). 

The authors thank A.\ M\'arffy for  his assistance with the pressure cell.

Samples were fabricated by B.\ Z., Z.\ K.\ K.\ and H.\ S.\ in G\"oteborg, Budapest and Manchester. Measurements were performed by T.\ P.\ and Z.\ K.\ K.\ in Budapest, and H.\ S.\ in Manchester. Analysis was carried out by T.\ P., M.\ L.\ M.\ and E.\ T.\ with the contribution of M.\ H.\ D.\ G. Research was guided by B.\ F., I.\ J.\ V.\ M., S.\ P.\ D., S.\ C., P.\ M.\ and E.\ T. 

\section*{Statements}\label{sec:con}

All data that support the findings of this study are included within the article (and any supplementary files).

The authors declare that no competing interests exist.

No human participants were included in this study besides the authors.

\FloatBarrier 

\bibliography{refs}

@book{Aharoni2020,
   author = {Amikan Aharoni},
   doi = {10.1093/oso/9780198508083.001.0001},
   isbn = {978-0198508090},
   publisher = {Oxford University Press},
   title = {An Introduction to the Theory of Ferromagnetism},
   year = {2000}
}

@article{Baltz2018,
   author = {V Baltz and A Manchon and M Tsoi and T Moriyama and T Ono and Y Tserkovnyak},
   doi = {10.1103/RevModPhys.90.015005},
   issue = {1},
   journal = {Reviews of Modern Physics},
   month = {2},
   pages = {15005},
   publisher = {American Physical Society},
   title = {Antiferromagnetic spintronics},
   volume = {90},
   url = {https://link.aps.org/doi/10.1103/RevModPhys.90.015005},
   year = {2018}
}

@book{Coey2010,
   author = {J. M. D. Coey},
   isbn = {9780511845000},
   pages = {195-199},
   publisher = {Cambridge University Press},
   title = {Magnetism and Magnetic Materials},
   year = {2010}
}

@article{2024Din,
   author = {A Dal Din and O J Amin and P Wadley and K W Edmonds},
   doi = {10.1038/s44306-024-00029-0},
   issn = {2948-2119},
   issue = {1},
   journal = {npj Spintronics},
   pages = {25},
   title = {Antiferromagnetic spintronics and beyond},
   volume = {2},
   url = {https://doi.org/10.1038/s44306-024-00029-0},
   year = {2024}
}

@article{Deng2018FGT,
   author = {Yujun Deng and Yijun Yu and Yichen Song and Jingzhao Zhang and Nai Zhou Wang and Zeyuan Sun and Yangfan Yi and Yi Zheng Wu and Shiwei Wu and Junyi Zhu and Jing Wang and Xian Hui Chen and Yuanbo Zhang},
   doi = {10.1038/s41586-018-0626-9},
   issn = {1476-4687},
   issue = {7729},
   journal = {Nature},
   pages = {94-99},
   title = {Gate-tunable room-temperature ferromagnetism in two-dimensional Fe3GeTe2},
   volume = {563},
   url = {https://doi.org/10.1038/s41586-018-0626-9},
   year = {2018}
}

@article{Elahi2022,
   author = {Ehsan Elahi and Ghulam Dastgeer and Ghazanfar Nazir and Sobia Nisar and Mudasar Bashir and Haroon Akhter Qureshi and Deok-kee Kim and Jamal Aziz and Muhammad Aslam and Kashif Hussain and Mohammed A Assiri and Muhammad Imran},
   doi = {https://doi.org/10.1016/j.commatsci.2022.111670},
   issn = {0927-0256},
   journal = {Computational Materials Science},
   pages = {111670},
   title = {A review on two-dimensional (2D) magnetic materials and their potential applications in spintronics and spin-caloritronic},
   volume = {213},
   url = {https://www.sciencedirect.com/science/article/pii/S0927025622004013},
   year = {2022}
}

@article{Ershadrad2022,
   author = {Soheil Ershadrad and Sukanya Ghosh and Duo Wang and Yaroslav Kvashnin and Biplab Sanyal},
   doi = {10.1021/acs.jpclett.2c00692},
   issue = {22},
   journal = {The Journal of Physical Chemistry Letters},
   month = {6},
   pages = {4877-4883},
   publisher = {American Chemical Society},
   title = {Unusual Magnetic Features in Two-Dimensional Fe5GeTe2 Induced by Structural Reconstructions},
   volume = {13},
   url = {https://doi.org/10.1021/acs.jpclett.2c00692},
   year = {2022}
}

@article{Feder1968,
   author = {J Feder and E Pytte},
   doi = {10.1103/PhysRev.168.640},
   issue = {2},
   journal = {Physical Review},
   month = {4},
   pages = {640-654},
   publisher = {American Physical Society},
   title = {Low-Temperature Behavior of the Anisotropic Heisenberg Antiferromagnet in the Neighborhood of the Magnetic Phase Boundaries},
   volume = {168},
   url = {https://link.aps.org/doi/10.1103/PhysRev.168.640},
   year = {1968}
}

@article{Fei2018,
   author = {Zaiyao Fei and Bevin Huang and Paul Malinowski and Wenbo Wang and Tiancheng Song and Joshua Sanchez and Wang Yao and Di Xiao and Xiaoyang Zhu and Andrew F May and Weida Wu and David H Cobden and Jiun-Haw Chu and Xiaodong Xu},
   doi = {10.1038/s41563-018-0149-7},
   issn = {1476-4660},
   issue = {9},
   journal = {Nature Materials},
   pages = {778-782},
   title = {Two-dimensional itinerant ferromagnetism in atomically thin Fe3GeTe2},
   volume = {17},
   url = {https://doi.org/10.1038/s41563-018-0149-7},
   year = {2018}
}

@article{Fulop2021,
   author = {Bálint Fülöp and Albin Márffy and Endre Tóvári and Máté Kedves and Simon Zihlmann and David Indolese and Zoltán Kovács-Krausz and Kenji Watanabe and Takashi Taniguchi and Christian Schönenberger and István Kézsmárki and Péter Makk and Szabolcs Csonka},
   doi = {10.1063/5.0058583},
   issn = {0021-8979},
   issue = {6},
   journal = {Journal of Applied Physics},
   month = {8},
   pages = {064303},
   publisher = {AIP Publishing LLCAIP Publishing},
   title = {New method of transport measurements on van der Waals heterostructures under pressure},
   volume = {130},
   url = {https://aip.scitation.org/doi/abs/10.1063/5.0058583},
   year = {2021}
}

@article{Jia2025,
   author = {Zhiyan Jia and Mengfan Zhao and Qian Chen and Yuxin Tian and Lixuan Liu and Fang Zhang and Delin Zhang and Yue Ji and Bruno Camargo and Kun Ye and Rong Sun and Zhongchang Wang and Yong Jiang},
   doi = {10.1021/acsnano.4c14168},
   issn = {1936-0851},
   issue = {10},
   journal = {ACS Nano},
   month = {3},
   note = {doi: 10.1021/acsnano.4c14168},
   pages = {9452-9483},
   publisher = {American Chemical Society},
   title = {Spintronic Devices upon 2D Magnetic Materials and Heterojunctions},
   volume = {19},
   url = {https://doi.org/10.1021/acsnano.4c14168},
   year = {2025}
}

@article{Kim2019,
   author = {Dongseuk Kim and Sijin Park and Jinhwan Lee and Jungbum Yoon and Sungjung Joo and Taeyueb Kim and Kil-joon Min and Seung-Young Park and Changsoo Kim and Kyoung-Woong Moon and Changgu Lee and Jisang Hong and Chanyong Hwang},
   doi = {10.1088/1361-6528/ab0a37},
   issn = {0957-4484},
   issue = {24},
   journal = {Nanotechnology},
   pages = {245701},
   publisher = {IOP Publishing},
   title = {Antiferromagnetic coupling of van der Waals ferromagnetic Fe3GeTe2},
   volume = {30},
   url = {https://doi.org/10.1088/1361-6528/ab0a37},
   year = {2019}
}

@article{Klein2018,
   author = {D R Klein and D Macneill and J L Lado and D Soriano and E Navarro-Moratalla and K Watanabe and T Taniguchi and S Manni and P Canfield and J Fernández-Rossier and P Jarillo-Herrero},
   doi = {10.1126/science.aar3617},
   journal = {Science},
   pages = {1218},
   title = {Probing magnetism in 2D van der Waals crystalline insulators via electron tunneling},
   volume = {360},
   url = {https://www.science.org/doi/10.1126/science.aar3617},
   year = {2018}
}

@article{Li2019NatMat,
   author = {Tingxin Li and Shengwei Jiang and Nikhil Sivadas and Zefang Wang and Yang Xu and Daniel Weber and Joshua E Goldberger and Kenji Watanabe and Takashi Taniguchi and Craig J Fennie and Kin Fai Mak and Jie Shan},
   doi = {10.1038/s41563-019-0506-1},
   issn = {1476-4660},
   issue = {12},
   journal = {Nature Materials},
   pages = {1303-1308},
   title = {Pressure-controlled interlayer magnetism in atomically thin CrI3},
   volume = {18},
   url = {https://doi.org/10.1038/s41563-019-0506-1},
   year = {2019}
}

@article{Li2023,
   author = {Yue Li and Xiaobing Hu and Arash Fereidouni and Rabindra Basnet and Krishna Pandey and Jianguo Wen and Yuzi Liu and Hong Zheng and Hugh O H Churchill and Jin Hu and Amanda K Petford-Long and Charudatta Phatak},
   doi = {10.1021/acsanm.2c05479},
   issue = {6},
   journal = {ACS Applied Nano Materials},
   month = {3},
   pages = {4390-4397},
   publisher = {American Chemical Society},
   title = {Visualizing the Effect of Oxidation on Magnetic Domain Behavior of Nanoscale Fe3GeTe2 for Applications in Spintronics},
   volume = {6},
   url = {https://doi.org/10.1021/acsanm.2c05479},
   year = {2023}
}

@article{Lu2024NanoLett,
   author = {Longyu Lu and Qing Wang and Hengli Duan and Kejia Zhu and Tao Hu and Yupeng Ma and Shengchun Shen and Yuran Niu and Jiatu Liu and Jianlin Wang and Sandy Adhitia Ekahana and Jan Dreiser and Y Soh and Wensheng Yan and Guopeng Wang and Yimin Xiong and Ning Hao and Yalin Lu and Mingliang Tian},
   doi = {10.1021/acs.nanolett.4c00472},
   issn = {1530-6984},
   issue = {20},
   journal = {Nano Letters},
   month = {5},
   note = {doi: 10.1021/acs.nanolett.4c00472},
   pages = {5984-5992},
   publisher = {American Chemical Society},
   title = {Tunable Magnetism in Atomically Thin Itinerant Antiferromagnet with Room-Temperature Ferromagnetic Order},
   volume = {24},
   url = {https://doi.org/10.1021/acs.nanolett.4c00472},
   year = {2024}
}

@article{May2019_F5GT,
   author = {Andrew F May and Dmitry Ovchinnikov and Qiang Zheng and Raphael Hermann and Stuart Calder and Bevin Huang and Zaiyao Fei and Yaohua Liu and Xiaodong Xu and Michael A McGuire},
   doi = {10.1021/acsnano.8b09660},
   issn = {1936-0851},
   issue = {4},
   journal = {ACS Nano},
   month = {4},
   note = {doi: 10.1021/acsnano.8b09660},
   pages = {4436-4442},
   publisher = {American Chemical Society},
   title = {Ferromagnetism Near Room Temperature in the Cleavable van der Waals Crystal Fe5GeTe2},
   volume = {13},
   url = {https://doi.org/10.1021/acsnano.8b09660},
   year = {2019}
}

@article{May2020PRM,
   author = {Andrew F May and Mao-Hua Du and Valentino R Cooper and Michael A McGuire},
   doi = {10.1103/PhysRevMaterials.4.074008},
   issue = {7},
   journal = {Physical Review Materials},
   month = {7},
   pages = {74008},
   publisher = {American Physical Society},
   title = {Tuning magnetic order in the van der Waals metal Fe5GeTe2 by cobalt substitution},
   volume = {4},
   url = {https://link.aps.org/doi/10.1103/PhysRevMaterials.4.074008},
   year = {2020}
}

@article{Mi2023,
   author = {Mengjuan Mi and Han Xiao and Lixuan Yu and Yingxu Zhang and Yuanshuo Wang and Qiang Cao and Yilin Wang},
   doi = {https://doi.org/10.1016/j.mtnano.2023.100408},
   issn = {2588-8420},
   journal = {Materials Today Nano},
   pages = {100408},
   title = {Two-dimensional magnetic materials for spintronic devices},
   volume = {24},
   url = {https://www.sciencedirect.com/science/article/pii/S2588842023001074},
   year = {2023}
}

@article{Marffy2026,
   author = {Albin Márffy and Endre Tóvári and Yu-Fei Liu and Anyuan Gao and Tianye Huang and László Oroszlány and Kenji Watanabe and Takashi Taniguchi and Su-Yang Xu and Péter Makk and Szabolcs Csonka},
   doi = {10.1021/acs.nanolett.5c05229},
   issn = {1530-6984},
   issue = {5},
   journal = {Nano Letters},
   month = {2},
   note = {doi: 10.1021/acs.nanolett.5c05229},
   pages = {1782-1788},
   publisher = {American Chemical Society},
   title = {Pressure-Tunable Phase Transitions in Atomically Thin Chern Insulator MnBi2Te4},
   volume = {26},
   url = {https://doi.org/10.1021/acs.nanolett.5c05229},
   year = {2026}
}

@article{Nagaosa2010,
   author = {Naoto Nagaosa and Jairo Sinova and Shigeki Onoda and A H MacDonald and N P Ong},
   doi = {10.1103/RevModPhys.82.1539},
   issue = {2},
   journal = {Reviews of Modern Physics},
   month = {5},
   pages = {1539-1592},
   publisher = {American Physical Society},
   title = {Anomalous Hall effect},
   volume = {82},
   url = {https://link.aps.org/doi/10.1103/RevModPhys.82.1539},
   year = {2010}
}

@article{Ngaloy2024,
   author = {Roselle Ngaloy and Bing Zhao and Soheil Ershadrad and Rahul Gupta and Masoumeh Davoudiniya and Lakhan Bainsla and Lars Sjöström and Md. Anamul Hoque and Alexei Kalaboukhov and Peter Svedlindh and Biplab Sanyal and Saroj Prasad Dash},
   doi = {10.1021/acsnano.3c07462},
   issn = {1936-0851},
   issue = {7},
   journal = {ACS Nano},
   month = {2},
   pages = {5240-5248},
   publisher = {American Chemical Society},
   title = {Strong In-Plane Magnetization and Spin Polarization in (Co0.15Fe0.85)5GeTe2/Graphene van der Waals Heterostructure Spin-Valve at Room Temperature},
   volume = {18},
   url = {https://doi.org/10.1021/acsnano.3c07462},
   year = {2024}
}

@article{Ningrum2020,
   author = {Vertikasari P Ningrum and Bowen Liu and Wei Wang and Yao Yin and Yi Cao and Chenyang Zha and Hongguang Xie and Xiaohong Jiang and Yan Sun and Sichen Qin and Xiaolong Chen and Tianshi Qin and Chao Zhu and Lin Wang and Wei Huang},
   doi = {10.34133/2020/1768918},
   journal = {Research},
   month = {3},
   note = {doi: 10.34133/2020/1768918},
   pages = {1768918},
   publisher = {American Association for the Advancement of Science},
   title = {Recent Advances in Two-Dimensional Magnets: Physics and Devices towards Spintronic Applications},
   url = {https://doi.org/10.34133/2020/1768918},
   year = {2020}
}

@article{Pellet2025,
   author = {Clément Pellet-Mary and Debarghya Dutta and Märta A Tschudin and Patrick Siegwolf and Boris Gross and David A Broadway and Jordan Cox and Carolin Schrader and Jodok Happacher and Daniel G Chica and Cory R Dean and Xavier Roy and Patrick Maletinsky},
   doi = {10.1038/s41467-025-64700-8},
   issn = {2041-1723},
   issue = {1},
   journal = {Nature Communications},
   pages = {9725},
   title = {Lateral exchange bias for Néel-vector control in atomically thin antiferromagnets},
   volume = {16},
   url = {https://doi.org/10.1038/s41467-025-64700-8},
   year = {2025}
}

@article{Roche2024,
   author = {Stephan Roche and Bart van Wees and Kevin Garello and Sergio O Valenzuela},
   doi = {10.1088/2053-1583/ad64e2},
   issn = {2053-1583},
   issue = {4},
   journal = {2D Materials},
   pages = {43001},
   publisher = {IOP Publishing},
   title = {Spintronics with two-dimensional materials and van der Waals heterostructures},
   volume = {11},
   url = {https://doi.org/10.1088/2053-1583/ad64e2},
   year = {2024}
}

@article{Roemer2020,
   author = {Ryan Roemer and Chong Liu and Ke Zou},
   doi = {10.1038/s41699-020-00167-z},
   issn = {2397-7132},
   issue = {1},
   journal = {npj 2D Materials and Applications},
   pages = {33},
   title = {Robust ferromagnetism in wafer-scale monolayer and multilayer Fe3GeTe2},
   volume = {4},
   url = {https://doi.org/10.1038/s41699-020-00167-z},
   year = {2020}
}

@article{Sierra2021,
   author = {Juan F Sierra and Jaroslav Fabian and Roland K Kawakami and Stephan Roche and Sergio O Valenzuela},
   doi = {10.1038/s41565-021-00936-x},
   issn = {1748-3395},
   issue = {8},
   journal = {Nature Nanotechnology},
   pages = {856-868},
   title = {Van der Waals heterostructures for spintronics and opto-spintronics},
   volume = {16},
   url = {https://doi.org/10.1038/s41565-021-00936-x},
   year = {2021}
}

@article{Song2019NatMat,
   author = {Tiancheng Song and Zaiyao Fei and Matthew Yankowitz and Zhong Lin and Qianni Jiang and Kyle Hwangbo and Qi Zhang and Bosong Sun and Takashi Taniguchi and Kenji Watanabe and Michael A McGuire and David Graf and Ting Cao and Jiun-Haw Chu and David H Cobden and Cory R Dean and Di Xiao and Xiaodong Xu},
   issn = {1476-4660},
   issue = {12},
   journal = {Nature Materials},
   month = {12},
   pages = {1298-1302},
   title = {Switching 2D magnetic states via pressure tuning of layer stacking},
   volume = {18},
   url = {https://doi.org/10.1038/s41563-019-0505-2},
   year = {2019}
}

@article{Song2018,
   author = {Tiancheng Song and Xinghan Cai and Matisse Wei-Yuan Tu and Xiaoou Zhang and Bevin Huang and Nathan P Wilson and Kyle L Seyler and Lin Zhu and Takashi Taniguchi and Kenji Watanabe and Michael A McGuire and David H Cobden and Di Xiao and Wang Yao and Xiaodong Xu},
   doi = {10.1126/science.aar4851},
   journal = {Science},
   pages = {1214},
   title = {Giant tunneling magnetoresistance in spin-filter van der Waals heterostructures},
   volume = {360},
   url = {https://www.science.org/doi/10.1126/science.aar4851},
   year = {2018}
}

@article{Song2023,
   author = {Jizhe Song and Jianing Chen and Mengtao Sun},
   doi = {https://doi.org/10.1016/j.mtelec.2023.100070},
   issn = {2772-9494},
   journal = {Materials Today Electronics},
   pages = {100070},
   title = {Van der Waals engineering toward designer spintronic heterostructures},
   volume = {6},
   url = {https://www.sciencedirect.com/science/article/pii/S2772949423000463},
   year = {2023}
}

@article{Tan2018,
   author = {Cheng Tan and Jinhwan Lee and Soon-Gil Jung and Tuson Park and Sultan Albarakati and James Partridge and Matthew R Field and Dougal G McCulloch and Lan Wang and Changgu Lee},
   doi = {10.1038/s41467-018-04018-w},
   issn = {2041-1723},
   issue = {1},
   journal = {Nature Communications},
   pages = {1554},
   title = {Hard magnetic properties in nanoflake van der Waals Fe3GeTe2},
   volume = {9},
   url = {https://doi.org/10.1038/s41467-018-04018-w},
   year = {2018}
}

@article{Thapa2025,
   author = {Abinash Thapa and Bikash Sharma},
   doi = {https://doi.org/10.1002/admt.202500133},
   issn = {2365-709X},
   issue = {16},
   journal = {Advanced Materials Technologies},
   month = {8},
   pages = {e00133},
   publisher = {John Wiley \& Sons, Ltd},
   title = {A Review on Novel Low-Dimensional Materials based Magnetic Tunnel Junctions: Opportunities, Challenges, and Applications},
   volume = {10},
   url = {https://doi.org/10.1002/admt.202500133},
   year = {2025}
}

@article{Tian2020APL,
   author = {Congkuan Tian and Feihao Pan and Sheng Xu and Kun Ai and Tianlong Xia and Peng Cheng},
   doi = {10.1063/5.0006337},
   issn = {0003-6951},
   issue = {20},
   journal = {Applied Physics Letters},
   month = {5},
   pages = {202402},
   title = {Tunable magnetic properties in van der Waals crystals (Fe1-xCox)5GeTe2},
   volume = {116},
   url = {https://doi.org/10.1063/5.0006337},
   year = {2020}
}

@article{Tschudin2024,
   author = {Märta A Tschudin and David A Broadway and Patrick Siegwolf and Carolin Schrader and Evan J Telford and Boris Gross and Jordan Cox and Adrien E E Dubois and Daniel G Chica and Ricardo Rama-Eiroa and Elton J. G. Santos and Martino Poggio and Michael E Ziebel and Cory R Dean and Xavier Roy and Patrick Maletinsky},
   doi = {10.1038/s41467-024-49717-9},
   issn = {2041-1723},
   issue = {1},
   journal = {Nature Communications},
   pages = {6005},
   title = {Imaging nanomagnetism and magnetic phase transitions in atomically thin CrSBr},
   volume = {15},
   url = {https://doi.org/10.1038/s41467-024-49717-9},
   year = {2024}
}

@article{Tyson2025,
   author = {Trevor A Tyson and Sandun Amarasinghe and A M Milinda Abeykoon and Roger Lalancette and Kai Du and Xiaochen Fang and Sang-W Cheong and Abdullah Al-Mahboob and Jerzy T Sadowski},
   doi = {10.1088/2053-1583/adb43f},
   issn = {2053-1583},
   issue = {2},
   journal = {2D Materials},
   pages = {025021},
   publisher = {IOP Publishing},
   title = {Surface magnetism in Fe3GeTe2 van der Waals ferromagnet},
   volume = {12},
   url = {https://doi.org/10.1088/2053-1583/adb43f},
   year = {2025}
}

@article{Wang2019,
   author = {Zhe Wang and Marco Gibertini and Dumitru Dumcenco and Takashi Taniguchi and Kenji Watanabe and Enrico Giannini and Alberto F Morpurgo},
   doi = {10.1038/s41565-019-0565-0},
   issn = {1748-3395},
   issue = {12},
   journal = {Nature Nanotechnology},
   pages = {1116-1122},
   title = {Determining the phase diagram of atomically thin layered antiferromagnet CrCl3},
   volume = {14},
   url = {https://doi.org/10.1038/s41565-019-0565-0},
   year = {2019}
}

@article{Wang2024b,
   author = {Mingjie Wang and Bin Lei and Kejia Zhu and Yazhou Deng and Mingliang Tian and Ziji Xiang and Tao Wu and Xianhui Chen},
   doi = {10.1038/s41699-024-00460-1},
   issn = {2397-7132},
   issue = {1},
   journal = {npj 2D Materials and Applications},
   pages = {22},
   title = {Hard ferromagnetism in van der Waals Fe3GaTe2 nanoflake down to monolayer},
   volume = {8},
   url = {https://doi.org/10.1038/s41699-024-00460-1},
   year = {2024}
}

@article{Yang2022,
   author = {Hyunsoo Yang and Sergio O Valenzuela and Mairbek Chshiev and Sébastien Couet and Bernard Dieny and Bruno Dlubak and Albert Fert and Kevin Garello and Matthieu Jamet and Dae-Eun Jeong and Kangho Lee and Taeyoung Lee and Marie-Blandine Martin and Gouri Sankar Kar and Pierre Sénéor and Hyeon-Jin Shin and Stephan Roche},
   doi = {10.1038/s41586-022-04768-0},
   issn = {1476-4687},
   issue = {7915},
   journal = {Nature},
   pages = {663-673},
   title = {Two-dimensional materials prospects for non-volatile spintronic memories},
   volume = {606},
   url = {https://doi.org/10.1038/s41586-022-04768-0},
   year = {2022}
}

@article{Ye2022,
   author = {Chen Ye and Cong Wang and Qiong Wu and Sheng Liu and Jiayuan Zhou and Guopeng Wang and Aljoscha Söll and Zdenek Sofer and Ming Yue and Xue Liu and Mingliang Tian and Qihua Xiong and Wei Ji and Xiao Renshaw Wang},
   doi = {10.1021/acsnano.2c01151},
   issn = {1936-0851},
   issue = {8},
   journal = {ACS Nano},
   month = {8},
   note = {doi: 10.1021/acsnano.2c01151},
   pages = {11876-11883},
   publisher = {American Chemical Society},
   title = {Layer-Dependent Interlayer Antiferromagnetic Spin Reorientation in Air-Stable Semiconductor CrSBr},
   volume = {16},
   url = {https://doi.org/10.1021/acsnano.2c01151},
   year = {2022}
}

@article{Zhang2022_FGaT,
   author = {Gaojie Zhang and Fei Guo and Hao Wu and Xiaokun Wen and Li Yang and Wen Jin and Wenfeng Zhang and Haixin Chang},
   doi = {10.1038/s41467-022-32605-5},
   issn = {2041-1723},
   issue = {1},
   journal = {Nature Communications},
   pages = {5067},
   title = {Above-room-temperature strong intrinsic ferromagnetism in 2D van der Waals Fe3GaTe2 with large perpendicular magnetic anisotropy},
   volume = {13},
   url = {https://doi.org/10.1038/s41467-022-32605-5},
   year = {2022}
}

@article{Zhang2022PRM,
   author = {Hongrui Zhang and Yu-Tsun Shao and Rui Chen and Xiang Chen and Sandhya Susarla and David Raftrey and Jonathan T Reichanadter and Lucas Caretta and Xiaoxi Huang and Nicholas S Settineri and Zhen Chen and Jingcheng Zhou and Edith Bourret-Courchesne and Peter Ercius and Jie Yao and Peter Fischer and Jeffrey B Neaton and David A Muller and Robert J Birgeneau and Ramamoorthy Ramesh},
   doi = {10.1103/PhysRevMaterials.6.044403},
   issue = {4},
   journal = {Physical Review Materials},
   month = {4},
   pages = {44403},
   publisher = {American Physical Society},
   title = {A room temperature polar magnetic metal},
   volume = {6},
   url = {https://link.aps.org/doi/10.1103/PhysRevMaterials.6.044403},
   year = {2022}
}

@article{Zhang2024AdvMat,
   author = {Hongrui Zhang and Xiang Chen and Tianye Wang and Xiaoxi Huang and Xianzhe Chen and Yu-Tsun Shao and Fanhao Meng and Peter Meisenheimer and Alpha N'Diaye and Christoph Klewe and Padraic Shafer and Hao Pan and Yanli Jia and Michael F Crommie and Lane W Martin and Jie Yao and Ziqiang Qiu and David A Muller and Robert J Birgeneau and Ramamoorthy Ramesh},
   doi = {https://doi.org/10.1002/adma.202308555},
   issn = {0935-9648},
   issue = {9},
   journal = {Advanced Materials},
   month = {3},
   pages = {2308555},
   publisher = {John Wiley \& Sons, Ltd},
   title = {Room-Temperature, Current-Induced Magnetization Self-Switching in A Van Der Waals Ferromagnet},
   volume = {36},
   url = {https://doi.org/10.1002/adma.202308555},
   year = {2024}
}

@article{Zhao2025aAdvMat,
   author = {Bing Zhao and Lakhan Bainsla and Soheil Ershadrad and Lunjie Zeng and Roselle Ngaloy and Peter Svedlindh and Eva Olsson and Biplab Sanyal and Saroj P Dash},
   doi = {https://doi.org/10.1002/adma.202502822},
   issn = {0935-9648},
   issue = {37},
   journal = {Advanced Materials},
   month = {9},
   pages = {2502822},
   publisher = {John Wiley \& Sons, Ltd},
   title = {Coexisting Non-Trivial Van der Waals Magnetic Orders Enable Field-Free Spin-Orbit Torque Magnetization Dynamics},
   volume = {37},
   url = {https://doi.org/10.1002/adma.202502822},
   year = {2025}
}

@article{Zhao2026arxiv,
   author = {Bing Zhao and Roselle Ngaloy and Lalit Pandey and Himanshu Bangar and Divya P Dubey and Saroj P Dash},
   doi = {10.48550/arXiv.2602.24046},
   journal = {arXiv e-prints},
   month = {2},
   pages = {arXiv:2602.24046},
   title = {Nanoelectronics with Two Dimensional Magnets},
   year = {2026}
}

@article{Zhao2023AM,
   author = {Bing Zhao and Roselle Ngaloy and Sukanya Ghosh and Soheil Ershadrad and Rahul Gupta and Khadiza Ali and Anamul Md. Hoque and Bogdan Karpiak and Dmitrii Khokhriakov and Craig Polley and Balasubramanian Thiagarajan and Alexei Kalaboukhov and Peter Svedlindh and Biplab Sanyal and Saroj P Dash},
   doi = {https://doi.org/10.1002/adma.202209113},
   issn = {0935-9648},
   issue = {16},
   journal = {Advanced Materials},
   month = {4},
   pages = {2209113},
   publisher = {John Wiley \& Sons, Ltd},
   title = {A Room-Temperature Spin-Valve with van der Waals Ferromagnet Fe5GeTe2/Graphene Heterostructure},
   volume = {35},
   url = {https://doi.org/10.1002/adma.202209113},
   year = {2023}
}

\balancecolsandclearpage


\onecolumngrid
\fontsize{11}{12}\selectfont
\appendix*

\counterwithin*{figure}{part}
\stepcounter{part}
\renewcommand{\thefigure}{S\arabic{figure}}

\counterwithin*{equation}{part}
\stepcounter{part}
\renewcommand{\theequation}{S\arabic{equation}}

\section*{Supplementary information}

\subsection{Hall resistance of CFGT samples}\label{SIsubsec:data}

\begin{figure}[hp]
\begin{center}
	\includegraphics[width=\textwidth]{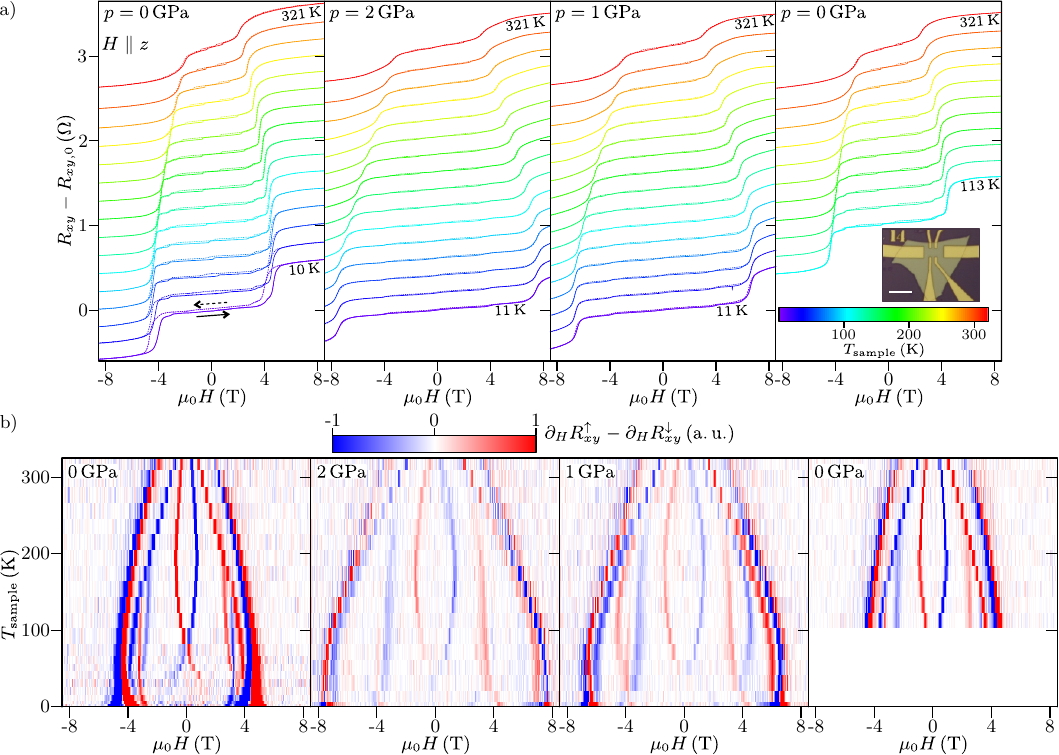}
	\caption{
    a) The Hall resistance of sample A ($90\,\mathrm{nm}$-thick) at different temperatures and pressures. The series of pressures are in the chronological order of measurements. The offset of each curve is proportional to the corresponding temperature. The inset is the optical image of the sample, the scale bar is $10\,\mathrm{\mu m}$. b) The difference of the $H$-derivatives of the Hall-resistances measured in up sweep and down sweep.
	}\label{Sfg:smplA}\end{center}
\end{figure}

\begin{figure}[hp]
\begin{center}
	\includegraphics[width=\textwidth]{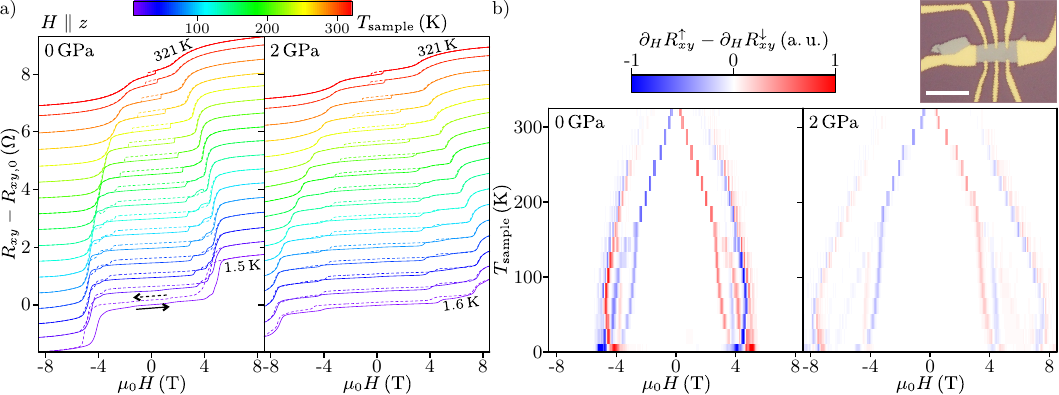}
	\caption{
    a) The Hall resistance of sample B ($33\,\mathrm{nm}$ thick) at different temperatures and pressures. The offset of the curves is proportional to the temperatures.
    b) The difference of the $H$-derivatives of the Hall-resistances measured in up sweep and down sweep. The inset is the optical image of the sample, the scale bar is $10\,\mathrm{\mu m}$.
	}\label{Sfg:smplB}\end{center}
\end{figure}

\begin{figure}[hp]
\begin{center}
	\includegraphics[width=\textwidth]{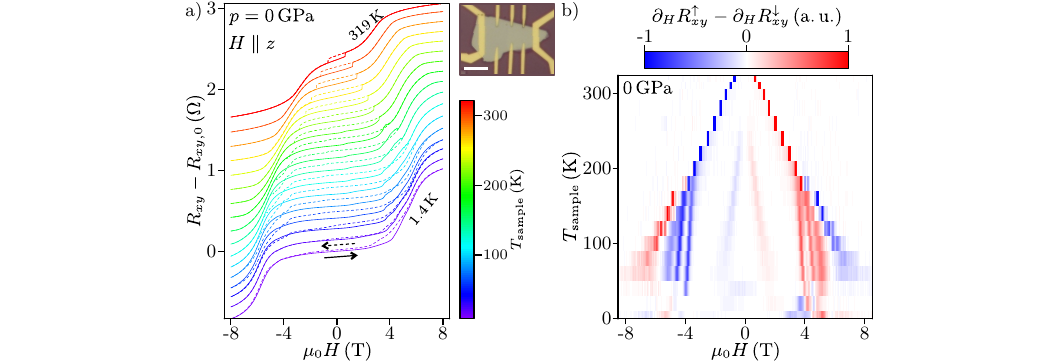}
	\caption{
    a) The Hall resistance of sample C ($43\,\mathrm{nm}$-thick) at 0 pressure. The offset of the curves is proportional to the temperatures. The inset shows the optical image of the sample, the scale bar is $10\,\mathrm{\mu m}$.
    b) The difference of the $H$-derivatives of the Hall-resistances measured in up sweep and down sweep.
	}\label{Sfg:smplC}\end{center}
\end{figure}

\begin{figure}[hp]
\begin{center}
	\includegraphics[width=\textwidth]{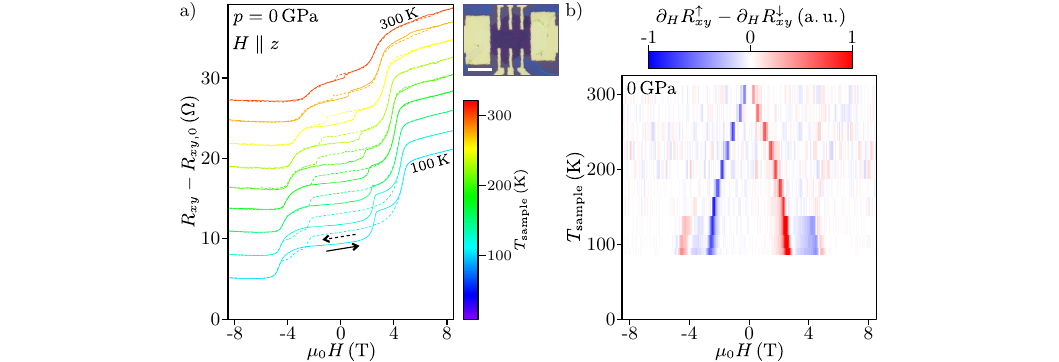}
	\caption{
    a) The Hall resistance of sample D (approximately $10.3\,\mathrm{nm}$-thick) at ambient pressure. The symmetric background is due to a mixing of the longitudinal resistance. The offset of the curves is proportional to the temperatures. The inset shows the optical image of the sample, the scale bar is $10\,\mathrm{\mu m}$.
    b) The difference of the $H$-derivatives of the Hall-resistances measured in up sweep and down sweep.
	}\label{Sfg:smplD}\end{center}
\end{figure}

Figures~\ref{Sfg:smplA}, \ref{Sfg:smplB}, \ref{Sfg:smplC} and \ref{Sfg:smplD} show measurements of the Hall resistance of samples A-D as a function of up and down field sweeps. In addition, the difference of the derivatives of the sweeps is plotted as colormaps as a function of $H$ and $T$, to better illustrate the evolution of hystereses with temperature. Especially, the low-$|H|$ coercive fields (inner edges) of the small loops in sample A are visible in the derivative map of the first panel of Fig.~\ref{Sfg:smplA}b as the red and blue ridges nearest $H=0$.

\begin{figure}[hp]
\begin{center}
	\includegraphics[width=\textwidth]{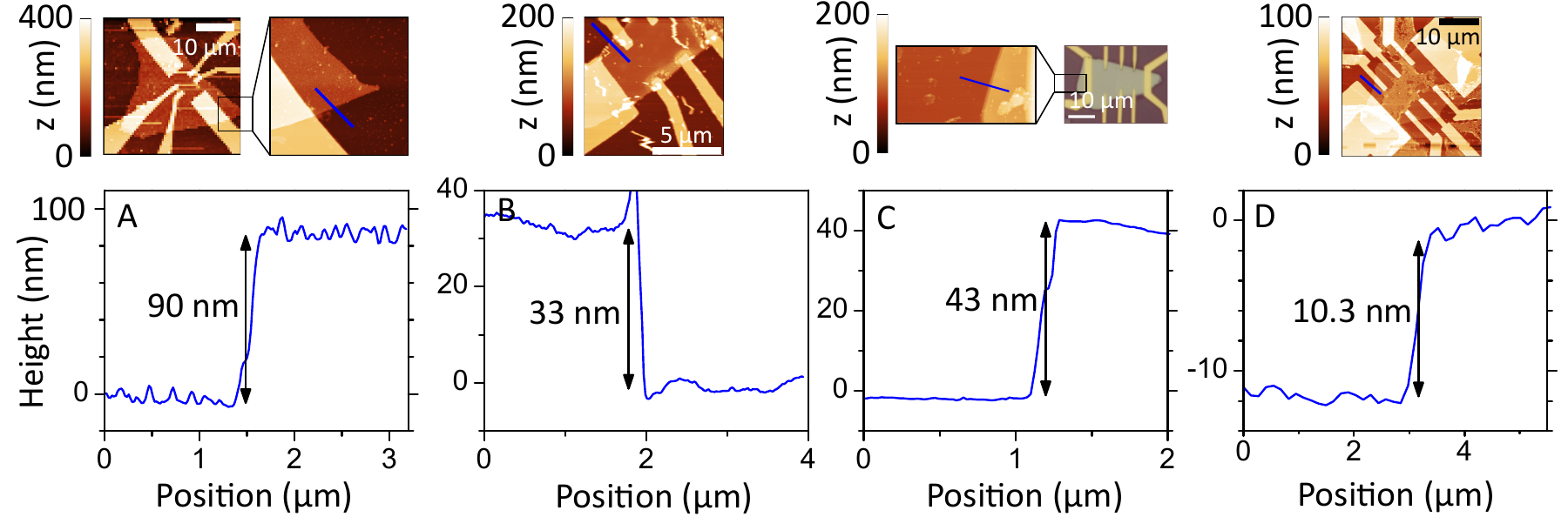}
	\caption{
    Atomic force microscopy images of the four samples. Blue lines in the colormaps represent the positions or the profiles shown in the graphs.
	}\label{Sfg:afm}\end{center}
\end{figure}

The thickness of the individual flakes was determined via atomic force microscopy. These measurements are shown in Fig.~\ref{Sfg:afm}.

\FloatBarrier

\subsection{Two-sublattice analysis}\label{SIsubsec:2sub}

\begin{figure}[hp]
\begin{center}
	\includegraphics[width=\textwidth]{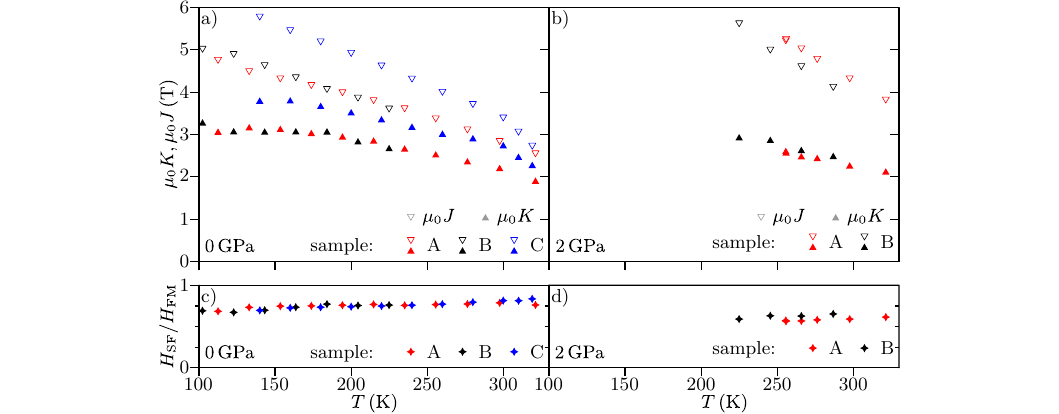}
	\caption{
    a-b) The comparison of $J$ and $K$ values (empty and full symbols, respectively) for the different samples at $0$ and $2\,\mathrm{GPa}$ pressures.
    c-d) $H_\mathrm{SF}/H_\mathrm{FM}$ as a function of temperature for the different samples at $0$ and $2\,\mathrm{GPa}$ pressures.
    }\label{Sfg:KJ}\end{center}
\end{figure}

We have performed the same two-sublattice model analysis for samples B, C as in the main text for sample A. The $T$-dependent $J,K$ parameters are plotted in Fig.~\ref{Sfg:KJ}a,b at 0 and 2\,GPa, respectively. The ratio of $H_\mathrm{SF}$ and $H_\mathrm{FM}$ is plotted in panels c,d. 

Regarding sample D, $H_\mathrm{FM}$ could not be resolved from the second derivative of $R_{xy}$. While $\Delta R$ could be estimated from the antisymmetrized data\cite{Marffy2026} for Fig.~4, this prevented the calculation of $J$ and $K$.

\newpage

\subsection{Global minimum calculations}\label{SIsubsec:sims}

\begin{figure}[hp]
\begin{center}
	\includegraphics[width=\textwidth]{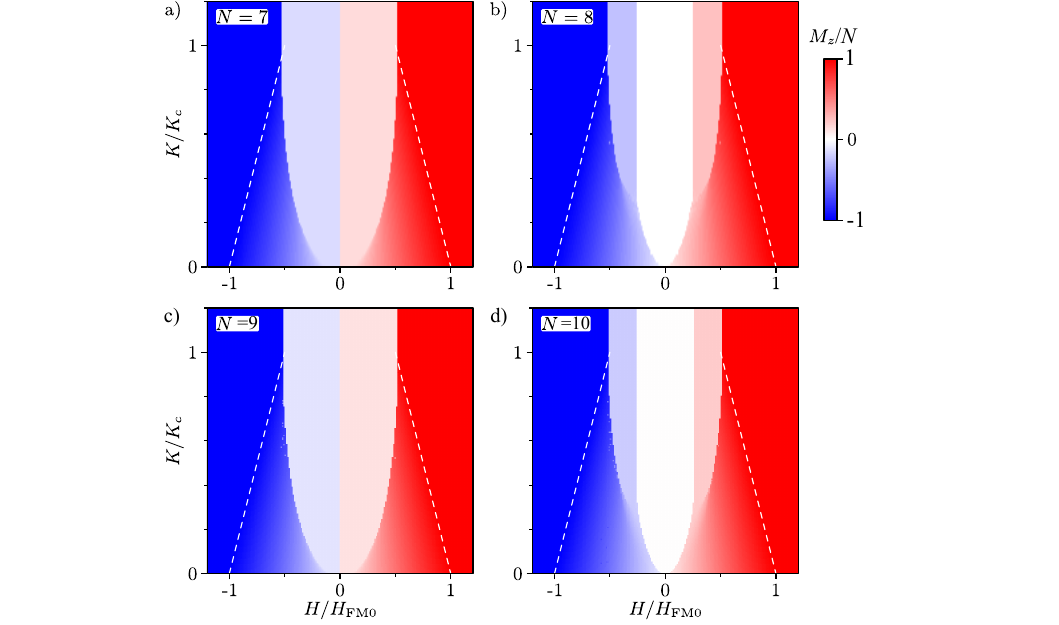}
	\caption{
    Numerical global minimum solutions of the linear chain model for a-d) $N = 7$ to $10$ layers. The horizontal scale is $H_\mathrm{FM0} = 2J \cos^2 \left( \pi/2N \right)$, which is the onset of the FM state for $K=0$. The vertical scale $K_c = H_\mathrm{FM0}-J$ is the largest $K$ where the cAFM state is a global minimum for $N>2$ \cite{Wang2019}. Dashed white lines indicate the border $H_\mathrm{FM} / H_\mathrm{FM0}$ between the cAFM and FM phases, which matches the numerical solution (colormap).}\label{Sfg:linch}\end{center}
\end{figure}

We have numerically calculated phase diagrams on the $H,K$ plane for $N=7$ to $10$ layers following the global minimum of the linear chain model in phase space, with $H$ applied along the easy axis. The colormaps in Fig.~\ref{Sfg:linch} show $M_z$ resulting from the calculations. Panels b,d) demonstrate for even number of layers the presence of intermediate states with magnetization $M_z=\pm2$ on either side of the AFM phase for large enough anisotropy $K$.

\FloatBarrier 

\end{document}